\documentclass[aps,10pt,prd,groupedaddress,nofootinbib,preprint]{revtex4-1}
\usepackage[utf8]{inputenc}
\usepackage{hyperref}
\usepackage{xcolor}
\usepackage{graphicx}
\usepackage{amsmath,amssymb}
\usepackage{bm}
\usepackage[utf8]{inputenc}

\begin{document}
	\preprint{YITP-16-113}
	\title{\boldmath Strongly scale-dependent CMB dipolar asymmetry \\ from super-curvature fluctuations}

	\author{\textsc{Christian Byrnes$^a$}}
	\email{C.Byrnes@sussex.ac.uk}
	
	\author{\textsc{Guillem Dom\`enech$^b$}}
	\email{guillem.domenech{}@{}yukawa.kyoto-u.ac.jp}
	
	\author{\textsc{Misao Sasaki$^b$} }
	\email{misao{}@{}yukawa.kyoto-u.ac.jp}
	
	\author{\textsc{Tomo Takahashi$^c$} }
	\email{tomot@cc.saga-u.ac.jp}
	
	\affiliation{
		$^{a}$\small{Department of Physics and  Astronomy, University of Sussex, Brighton BN1 9QH, UK}\\\\
		$^{b}$\small{Center for Gravitational Physics,\\Yukawa Institute for Theoretical Physics, Kyoto University, Kyoto 606-8502, Japan}\\\\
		$^{c}$\small{Department of Physics, Saga University, Saga 840-8502, Japan}
	}
	\date{\today}
	
	\begin{abstract} 
    We reconsider the observed CMB dipolar asymmetry in the context of open inflation, where a supercurvature mode might survive the bubble nucleation. If such a supercurvature mode modulates the amplitude of the curvature power spectrum, it would easily produce an asymmetry in the power spectrum. We show that current observational data can be accommodated in a three-field model, with simple quadratic potentials and a non-trivial field-space metric. Despite the presence of three fields, we believe this model is so far the simplest that can match current observations. We are able to match the observed strong scale dependence of the dipolar asymmetry, without a fine tuning of initial conditions, breaking slow roll or adding a feature to the evolution of any field.
	\end{abstract}

\maketitle
\section{Introduction}

It might well be that at very early times the universe was in a metastable vacuum state \cite{Coleman:1980aw}, e.g.~de Sitter false vacuum. Then, our observable universe would be a product of vacuum decay to a true vacuum bubble followed by a period of inflation \cite{Sasaki:1993gy,Bucher:1994gb}. If the subsequent inflation lasted just enough, i.e.~around $60$ $e$-folds, we may hope to detect signatures/effects of its origins, e.g.~in the large scales of the cosmic microwave background (CMB). For example, it could explain the large scale suppression observed in the CMB temperature power spectrum \cite{White:2014aua}. 

Of particular interest is the fact that recent observational data suggests the presence of a dipole anomaly at about $2\--3\,\sigma$ significance \cite{Eriksen:2003db,Hansen:2004vq,Eriksen:2007pc,Hoftuft:2009rq,Bennett:2010jb,Ade:2013nlj,Ade:2015hxq}. Despite not being overwhelmingly statistically significant, one cannot help being curious about understanding what could create such an anomaly. It becomes even more interesting when in a bubble universe our Hubble patch could be inside a much larger fluctuation \cite{1978SvA}, called a supercurvature mode \cite{Sasaki:1994yt,Lyth:1995cw,GarciaBellido:1995wz,GarciaBellido:1997te}. Such a mode could modulate the curvature power spectrum and, for example, it could generate multipole moments of the temperature fluctuations in the CMB \cite{Erickcek:2008sm,Erickcek:2008jp}. This mechanism to explain the dipolar asymmetry is so far the most pursued and promising candidate \cite{Erickcek:2008sm,Erickcek:2008jp,Erickcek:2009at,Kanno:2013ohv,Liddle:2013czu,Lyth:2013vha,Cai:2013gma,Assadullahi:2014pya,Byrnes:2015asa,Cai:2015xba,Byrnes:2015dub,Kobayashi:2015qma}. An open question of this model is whether such a supercurvature fluctuation was present before the tunneling and thus how it was generated or if it was the result of quasi-open inflation \cite{GarciaBellido:1997te}. Despite being an interesting issue, it is beyond the scope of this paper. Here we simply focus on the evolution of the supercurvature mode after the nucleation. For alternative scenarios see for example references \cite{D'Amico:2013iaa,Kohri:2013kqa,Jazayeri:2014nya,Ringeval:2015ywa,Firouzjahi:2016fxf}.



Although theoretically it is an attractive idea, in practice one finds severe constraints from CMB data. In fact, many models are ruled out when confronted with constraints of local type non-gaussianity and a quadrupolar modulation of the power spectrum \cite{Kanno:2013ohv,Byrnes:2015dub,Byrnes:2016zxb}. Furthermore, a recent analysis of the dipole asymmetry indicates a strong scale dependence \cite{Aiola:2015rqa} which complicates even more the discussion as one usually needs a large departure from slow-roll to explain such a scale dependence. Before entering the theoretical discussion, let us review the magnitude and scale dependence of current observational bounds. The departure from isotropy is parametrized as follows:
\begin{align}\label{eq:dipole}
	P(k)=P_{\rm iso}(k)\left(1+2A(k)\,\hat{\mathbf{n}}\cdot\hat{\mathbf{p}}+B(k)\left(\hat{\mathbf{n}}\cdot\hat{\mathbf{p}}\right)^2\right)\,,
\end{align}
where $\hat{\mathbf{n}}$ is a direction on the sky and $\hat{\mathbf{p}}$ is the direction of the dipole asymmetry. $A(k)$ and $B(k)$ are respectively the amplitude of the dipole and quadrupole modulation. 

Regarding the dipolar anomaly, it has been observed as $A (k) \sim 0.1$ on large scales from WMAP and Planck data  
\cite{Eriksen:2003db,Hansen:2004vq,Eriksen:2007pc,Hoftuft:2009rq,Ade:2013nlj,Ade:2015hxq}. 
On small scales, it has been severely constrained as \cite{Flender:2013jja} (see also \cite{Hirata:2009ar,Adhikari:2014mua,Quartin:2014yaa}) 
\begin{align}
A(k)< 0.0045,  \quad {\rm for}~{\rm \ell=601-2048} \quad (95 \%~ {\rm C.L.})\,,
\end{align}
which suggests that the dipolar asymmetry is strongly scale dependent. Such scale-dependence is studied in detail in  \cite{Aiola:2015rqa}, providing a fit to the Planck data in the multipole $\ell$ space as $A(\ell) \propto (\ell /60)^{n}$ with $n = -0.54^{+0.38}_{-0.22}$ (for $\ell < 200$), which corresponds to 68 \% C.L. For later reference, we roughly recast the current observational bound in $k$ space as
\begin{align}\label{eq:alarge}
	A(k)\approx 0.1 \,\left(\frac{k}{k_0}\right)^{-1/2}\,,
\end{align}
where $k_0$ is a pivot scale  given by $k_0 =10^{-4} ~ {\rm Mpc}^{-1}$.

As for the quadrupole, no direct constraint on $B(k)$ exists in the form defined by Eq.~\eqref{eq:dipole}. Still, we can deduce an upper bound from \cite{Kim:2013gka,Ade:2015ava} on a scale-invariant quadrupole modulation to be
\begin{align}\label{eq:blarge}
	|B| \lesssim 10^{-2}.
\end{align}
 Let us note that Kim and Komatsu \cite{Kim:2013gka} constrained the quadrupole asymmetry of an homogeneous but anisotropic universe, i.e.~with no dipole term. On the other hand, our parametrization corresponds to an inhomogeneous and anisotropic universe, due to the supercurvature mode. However, we can safely regard Eq.~\eqref{eq:blarge} as a conservative upper bound since, after decomposing the anisotropic part using spherical harmonics, the preferred direction only appears in the estimator.

In this paper, we discuss a model which predicts a
dipole asymmetry characterized by Eq.~\eqref{eq:alarge} and the quadrupole is bounded by Eq.~\eqref{eq:blarge}. We look for a mechanism that generates scale-dependent dipole asymmetry, but a negligible quadrupolar asymmetry. This work is organised as follows. In section \ref{sec:model} we review the model of Kanno, Sasaki and Tanaka \cite{Kanno:2013ohv} and confront it with current bounds. In section \ref{sec:example}, we give an explicit example that fits the observational results. We conclude our work in section \ref{sec:conclusions}.


\section{Model}\label{sec:model}

In this section we begin by briefly reviewing the model proposed by Kanno, Sasaki \& Tanaka \cite{Kanno:2013ohv}. The main idea is quite simple and it goes as follows. First, consider that there is a scalar field, let us say $\sigma$, with a supercurvature fluctuation, denoted by $\Delta\sigma$.
It is known that the $\sigma$ field should not be the field which drives inflation, otherwise we would either not see any effect or it would be ruled out by observations \cite{Erickcek:2008sm,Wang:2013lda}, nor be the field which generates the curvature perturbation, since in that case the quadrupole modulation is usually too large \cite{Kanno:2013ohv,Byrnes:2015asa}. As a result, we are left with two options. Either we consider a non-canonical mixed scalar field model as in \cite{Cai:2013gma,Cai:2015xba} or we add another scalar field \cite{Kanno:2013ohv}. In the latter case, it is fairly easy to build a model which satisfies the behaviour on large scales given by Eq.~\eqref{eq:alarge}.

The simplest three field model consists of one field $\phi$ that drives inflation,\footnote{Details of the inflationary dynamics are not relevant for our purposes.} one field $\chi$ that gives the dominant contribution to the curvature power spectrum through the curvaton mechanism \cite{Moroi:2002rd,Enqvist:2001zp,Lyth:2001nq} and the field $\sigma$ with a supercurvature mode $\Delta\sigma$ which couples to $\chi$ and modulates the spectrum. Concretely, the Lagrangian for the scalar fields is given by
\begin{align}\label{eq:lagrangian}
{\cal L}=-\frac{1}{2}(\nabla\phi)^2-V(\phi)-\frac{1}{2}f^2(\sigma)(\nabla\chi)^2-\frac{1}{2}m_\chi^2\chi^2-\frac{1}{2}(\nabla\sigma)^2-V(\sigma)\,.
\end{align}
The curvature power spectrum due to the curvaton is given by\footnote{It is easy to see that a factor $1/f^2(\sigma)$ appears due to normalization of the mode function for $\chi$.} 
\begin{align}\label{eq:power}
\mathcal{P}_{\zeta}=\frac{r_{\rm dec}^2}{9\pi^2}\frac{H^2}{\chi_*^2 f^2(\sigma_*)}\frac{(g')^2}{g^2}\,,
\end{align}
where $r_{\rm dec}$ roughly corresponds to the energy fraction of the curvaton at the time of its decay,  $\chi_\ast$ is its VEV at the time of horizon crossing, $g(\chi_\ast)$ denotes the initial amplitude of the curvaton oscillations in terms of $\chi_\ast$ and $g'  = dg /d\chi\big|_\ast$.
The local type non-gaussianity can be estimated using the $\delta N$ formalism as \cite{Sasaki:2006kq}
\begin{align}
	\label{eq:fNL_local}
	f_{\rm NL}^{\rm (local)} 
	= 
	\frac{5}{4r_{\rm dec}} \left( 1+ \frac{gg^{''}}{\left(g' \right)^2} \right) - \frac53 - \frac{5 r_{\rm dec}}{6}.
\end{align}
Attention should be paid since there is a mixing of the field $\chi$ with $\sigma$ through the kinetic term which could potentially modify the value of non-gaussianity, compared to the uncoupled case, through the equations of motion of $\chi$, i.e.~through $g(\chi_\ast)$. As will be pointed out below, $\sigma$ must be slowly rolling to satisfy the observational constraints and, therefore, we do not expect any significant modification at leading order in slow-roll. Moreover, we assumed that the $\sigma$ field does not contribute to the curvature perturbation and so any three-point correlation function involving the $\sigma$ field is negligible.

Under such assumptions, the local non-gaussianity $f^{\rm (local)}_{NL}$ essentially depends on the curvaton and can be made of order unity by choosing the curvaton parameters, e.g.~if the decay is late enough then $r_{\rm dec}=1$. It can be checked that when the curvaton potential is purely quadratic and $f(\sigma)=1$, then $ g{''}=0 $. Note that any constant $f(\sigma)=C$ can be absorbed into the mass of the curvaton. Next to leading order corrections to the non-Gaussianity may include scale dependence\footnote{
	The scale-dependence of $f_{\rm NL}$ generated from the evolution of the curvaton has been discussed in the 
	context of the self-interacting curvaton \cite{Byrnes:2009pe,Byrnes:2010xd,Byrnes:2011gh,Kobayashi:2012ba}.
}
and also a dipolar asymmetry due to the modulation by $\Delta\sigma$,  which in principle gives a distinctive signature of this model. For completeness, let us remind the reader that the latest constraint on local non-gaussianity combining temperature and polarization is given by $f_{NL}^{\rm (local)}=0.8\pm5.0$ ($68\%$ C.L.) \cite{Ade:2015ava}.

Before assuming any particular form of $f(\sigma)$ and $V(\sigma)$, let us take a look at the form of the dipole and quadrupole modulation. Expanding the curvaton power spectrum  \eqref{eq:power} around $\sigma$ yields
\begin{align}\label{eq:A}
  A(k)=\frac{\Delta {\cal P}_\chi}{2{\cal P}_\chi}=-\frac{\Delta\sigma}{\sigma}\frac{\sigma f_\sigma}{f} \,,
\end{align}
and  
\begin{align}\label{eq:B}
  B(k)=\frac{\Delta^{(2)} {\cal P}_\chi}{{\cal P}_\chi}=\left(	\frac{\Delta\sigma}{\sigma}\right)^2\left(3\frac{\sigma^2f^2_\sigma}{f^2}-\frac{\sigma^2f_{\sigma\sigma}}{f}\right)
  \,.
\end{align}
There is also a modulation in the spectral index given by  
\begin{align}
\label{eq:ns}
   n_s-1 &= -2\epsilon+2\frac{m_\chi^2}{3H^2}  -2 \frac{d\ln f}{d\ln k}=  
   -2\epsilon+2\frac{m_\chi^2}{3H^2} -2\frac{\sigma f_\sigma}{f}\frac{d\ln \sigma}{d\ln k} \nonumber
   \\ &=-2\epsilon+2\frac{m_\chi^2}{3H^2} -2A(k)\left(\frac{\Delta\sigma}{\sigma}\right)^{-1}\frac{d\ln \sigma}{d\ln k}\,,
\end{align}
and we used  Eq.~\eqref{eq:A} in the last step. Assuming that
\begin{align}\label{eq:amp}
\frac{\Delta\sigma}{\sigma}\sim 0.1\,,
\end{align}
we can infer from the amplitude of the dipole Eq.\eqref{eq:alarge} and the spectral index Eq.\eqref{eq:ns} that
\begin{align}\label{eq:slowroll}
	\frac{d\ln \sigma}{d\ln k}<10^{-2}\,,
\end{align}
which for a field value of $\sigma\sim M_{pl}$ is equivalent to the slow roll condition ${d(\sigma/M_{pl}})/{dN}<10^{-2}$. Recall that current observations suggest 
$n_s\sim 0.96$ \cite{Ade:2015lrj}. Also note that small-field models of inflation are unable to give a sufficiently red spectrum in this case because we require a dominant contribution to the spectral index from $\epsilon$, with quartic inflation providing a much better fit than other inflation models \cite{Enqvist:2013paa,Hardwick:2015tma,Vennin:2015vfa,Vennin:2015egh}. 

Let us turn now to the quadrupole modulation. For simplicity, let us assume that $f(\sigma)$ is given by a power-law, i.e.
\begin{align}\label{eq:f}
	f(\sigma)=\left(\frac{\sigma}{\sigma_0}\right)^\beta\,,
\end{align}
where $\sigma_0$ is some non-zero value of the field and $\beta$ is our free parameter.  Equations \eqref{eq:B} and \eqref{eq:amp} then suggest that any $|\beta|\sim O(1)$ would be enough to fit the quadrupole upper bound Eq.~\eqref{eq:blarge}, that is
\begin{align}
	& B(k)=\left(\frac{\Delta\sigma}{\sigma}\right)^2\beta\,(2\beta+1)=A^2(k)\,(2+{\beta}^{-1})\,.
\end{align}
where in the last step we used $A(k) = - \beta \frac{\Delta\sigma}{\sigma}$.
Note that with constant $f$, that is $\beta=0$, the asymmetry goes to zero. At this stage, the dipole asymmetry scale dependence is determined by the scale dependence of $\Delta\sigma$ (assuming that $\sigma$ is slowly rolling), which is essentially given in terms of the potential $V(\sigma)$. Let us show in the next section that a simple choice of $V(\sigma)$ could achieve the desired scale dependence Eq.~\eqref{eq:alarge}.

Before ending this section, let us note that the multipole moments $a_{i0}$ with $i=1,2,3$ are expected to be small and within the observational bounds quoted in \cite{Kanno:2013ohv,Byrnes:2015dub}. The reason is as follows: In the $\delta N$ formalism one has $\zeta\approx N_\chi \delta\chi+N_{\chi\chi}\delta\chi^2+...$, under the assumption that $\chi$ is the main contributor to the curvature perturbation. Since $\sigma$ does not affect the background dynamics much, $N_\chi$ and $N_{\chi\chi}$ only depend on the curvaton parameters and in particular non-gaussianity, corresponding to $N_{\chi\chi}$, is chosen to be small. Thus, the effect of $\sigma$ is a modulation on $\delta\chi$ (see Eq.\eqref{eq:power}), which is a higher-order correction to $a_{i0}$.

\section{A viable example}\label{sec:example}

In this section, we present a particular example which reproduces the desired scale dependence. Before using a detailed form of the potential let us mention some general assumptions and features. First, if the supercurvature mode is to survive the previous false vacuum de Sitter stage, the mass of the $\sigma$ field must be small compared to the Hubble parameter at that moment. Then, once the bubble is nucleated and the mode entered the true vacuum phase, the mass of $\sigma$ should be comparable to the true vacuum Hubble parameter so as to decay fast enough. Therefore, we assume that $V_{\sigma\sigma}\ll H_{False}^2$ before the tunneling and $V_{\sigma\sigma}\gtrsim H_{True}^2$ during the subsequent period of inflation. On the other hand, recall that  Eq.~\eqref{eq:slowroll} required $\sigma$ to be slowly rolling. Thus, we need a field that is massive and slowly rolling\footnote{We note that a slowly-rolling field can still have a large $\eta$ slow-roll parameter. In particular, for a spectator field it is possible for the hierarchy $\epsilon_\sigma=(V_\sigma/(3M_{Pl}H^2))^2/2\ll\eta_\sigma=V_{\sigma\sigma}/(3H^2)\simeq1$ to persist for a long time.}.

For the moment, let us consider the detailed evolution of $\sigma+\Delta\sigma$. Under the assumption that inflation is an almost de Sitter universe mainly driven by $\phi$, i.e. with an almost constant expansion factor $H$ or symbolically
\begin{align}
\epsilon\equiv\frac{-\dot H}{H^2}\ll 1\,,
\end{align}
the equations of motion for $\sigma$ and $\Delta\sigma$ from the Lagrangian \eqref{eq:lagrangian} are respectively given by
\begin{align}
&\frac{d\sigma/M_{pl}}{dN}=-\frac{V_\sigma}{3M_{pl}H^2} \,, \label{eq:1} \\ \medskip
&\frac{1}{3}\frac{d^2\Delta\sigma/M_{pl}}{dN^2}+\frac{d\Delta\sigma/M_{pl}}{dN}=\frac{-V_{\sigma\sigma}}{3H^2}\frac{\Delta\sigma}{M_{pl}}\label{eq:2}\,.
\end{align}
$N$ is the number of e-folds, that is $dN=Hdt$, and we neglected the coupling with the curvaton field as it is very light and slowly rolling. It should be noted that in  Eq.~\eqref{eq:1} we already used the slow roll assumption for $\sigma$.

Now, by requiring that $\Delta\sigma$ has a strong scale dependence with an approximately constant spectral index, i.e.
\begin{align}\label{eq:decay}
	\Delta\sigma= \Delta\sigma_0 \left(\frac{k}{k_0}\right)^{-\alpha}=\Delta\sigma_0\, {\rm e}^{-\alpha (N-N_0)}\,,
\end{align}
where we used the fact that at the time of horizon crossing $k=Ha$, which implies $d\ln k=d\ln a=dN$, one obtains from Eq.~\eqref{eq:2} that
\begin{align}\label{eq:vss}
	\frac{V_{\sigma\sigma}}{3H^2}=\alpha\left(1-\alpha/3\right)\,.
\end{align}
In our case of interest $\alpha\sim 1/2$ and $\eta_\sigma\equiv V_{\sigma\sigma}/(3H^2)=5/12$, see Eq.\,\eqref{eq:alarge}.  However, recall that the slow-roll condition Eq.~\eqref{eq:slowroll}  requires
\begin{align}
&\frac{d \sigma/M_{pl}}{dN}=-\frac{V_{\sigma}}{3M_{pl}H^2}<10^{-2}\label{eq:bg}\,.
\end{align}
These two requirements, equations \eqref{eq:vss} and \eqref{eq:bg}, imply that the field $\sigma$ has a mass of the order of $H$ and that after the bubble is nucleated $\sigma$ is placed very close to its minimum, say $\sigma_0$, so that its motion is negligible. The simplest example is a quadratic potential with its minimum at some $\sigma_0\sim M_{pl}$, i.e.
\begin{align}
	V(\sigma)=\frac{1}{2}m_\sigma^2\,\left({\sigma}-\sigma_0\right)^2\,.
\end{align}
More concretely, we require that the average value of the field in our observable patch is given by $\sigma\simeq\sigma_0$, while its value on opposite sides of the observed CMB in directions aligned to the asymmetry is $\sigma_0\pm\Delta\sigma/2$.

One may wonder for how many e-folds the classical description holds, since the $\sigma$ field is so close to its potential minimum that quantum diffusion could be relevant. However, note that it is the value of the field at the minimum of the potential, $\sigma_0$, which matters. One can easily realize this by doing a field redefinition $\tilde\sigma\equiv\sigma-\sigma_0$. In this case, the field rolls to $\tilde\sigma\to0$ and $f\propto1+\beta\,\tilde\sigma/\sigma_0$. Then, any small quantum fluctuation in $\tilde\sigma$ does not significantly affect the value of $f$, in fact it is suppressed by $\sigma_0\sim M_{pl}$ and $f_\sigma$ becomes a constant.
On the other hand, the evolution of $\Delta\sigma$ cannot be affected by small-scale quantum fluctuations since it is a super-horizon fluctuation, thus causally disconnected.

For example, taking $\Delta \sigma_0 / \sigma_0 = 0.1$ and $\beta = 1$, we obtain $A(k) = 0.1 (k/k_0)^{-1/2}$ and $B(k) = 0.03(k/k_0)^{-1}$, which are consistent with 
the current bounds. We note that taking a larger (or negative) value for $\beta$ while fixing  $\beta \left( \Delta \sigma_0 / \sigma_0 \right)= 0.1$ gives a smaller quadrupole $B$. Lastly, let us reiterate that unlike many previous models in the literature, non-Gaussianity will only be present at the level of $f_{\rm NL}\sim1$, only depending on the curvaton parameters.

\section{Conclusions}\label{sec:conclusions}


Anomalies in the CMB, such as the dipole asymmetry, have led to a long debate ranging from looking for any possible explanation to questioning the anomalies themselves. As far as the dipole asymmetry is concerned, it has persisted from WMAP \cite{Eriksen:2003db,Hansen:2004vq,Eriksen:2007pc,Hoftuft:2009rq,Bennett:2010jb} until Planck 2015 \cite{Ade:2013nlj,Ade:2015hxq}, currently with a statistical significance around $3\sigma$. Hopefully, precise analysis of CMB polarization data and future surveys of large scale structure, as well as their cross correlations, will help to tighten the bounds \cite{Chang:2013mya,Abolhasani:2013vaa,Fernandez-Cobos:2013fda,Hassani:2015zat,Shiraishi:2016omb,Yasini:2016pby}. On the other hand, the supercurvature mode explanation used in this and many works relies on open inflation, which could in principle be tested by future experiments on the spatial curvature of our universe \cite{Takada:2015mma}.

In this work we considered that the dipole asymmetry is generated in a three scalar field model \cite{Kanno:2013ohv}. One field drives inflation, one is responsible for the curvature perturbations through the curvaton mechanism and the latter is coupled to a third field which contains a supercurvature mode. Insensitive to the exact inflationary dynamics, we have shown that if the mass of the field $\sigma$ containing the supercurvature mode is approximately given by
\begin{align}
m^2_\sigma \sim H^2\,,
\end{align}
and the minimum of its potential is located at some non-zero value of the field, i.e. $\sigma_0\sim M_{\rm pl}$, the dipole asymmetry can be given by
\begin{align}
A(k)\sim-\beta \frac{\Delta\sigma_0}{\sigma_0}\,\left(\frac{k}{k_0}\right)^{-\alpha}\,,
\end{align}
where $\alpha=3/2-\sqrt{9/4-m_\sigma^2/H^2}$, $\Delta\sigma_0$ is the initial amplitude of the supercurvature mode and $\beta$ determines the coupling between $\sigma$ and the curvaton field $\chi$ whose explicit 
form is given in Eq.~\eqref{eq:f}. By appropriately choosing parameter values, we can obtain $A(k) \sim 0.1 (k/k_0)^{-1/2}$ as suggested by observations. 
A particular feature of this model is that, contrary to the usual two field case \cite{Kanno:2013ohv,Byrnes:2015asa,Byrnes:2015dub,Byrnes:2016zxb}, we can have a sizeable dipole asymmetry with a small quadrupole asymmetry, which is well within the bounds. 

It is interesting to note that more complex and finely tuned two-field models are also able to achieve such a strong scale dependence. For example, Ref.~\cite{Byrnes:2015dub} showed that a model where one field generates Gaussian scale-invariant perturbations and the second field generates the asymmetry and non-Gaussianity can only work with a lot of fine tuning and adding a feature to the potential (which implies that the strong scale dependence cannot persist for very long). However, the main point of this work is to show that the presence of a third field, with a mass of the order of $H$, considerably simplifies the model and easily reproduces a strong scale dependence as long as the VEV of such field is non-zero, e.g. $\sigma_0\sim M_{pl}$, passing current observational test. The only requirement is that the super curvature mode is sufficiently far displaced from its minimum, i.e. $\Delta\sigma_0/\sigma_0\sim0.1$ and that the two spectator fields are coupled through their kinetic terms. 

\section*{Acknowledgements}
This work was supported in part by MEXT KAKENHI Grant Number 15H05888, 15K21733 and 15K05084 (TT). CB is supported by a Royal Society University Research Fellowship. This work was initiated from discussions at the workshop ``inflationary universe", YITP-X-15-5, held at the Yukawa Institute for Theoretical Physics, Kyoto University. CB would like to thank the YITP for providing financial support to attend this meeting.

\bibliography{bibliography}

\end{document}